\documentclass{ws-procs9x6}

\usepackage{amssymb,epsfig}

\usepackage{psfrag}

\psfrag{kt}{{\footnotesize $k_\perp$}}
\psfrag{=}{{\footnotesize $=$}}
\psfrag{+}{{\footnotesize $+$}}
\psfrag{l}{{\footnotesize $l$}}
\psfrag{p1z'}{{\footnotesize $p_1z'$}}
\psfrag{q1}{{\footnotesize $q_1$}}
\psfrag{p1}{{\footnotesize $p_1$}}
\psfrag{-p1bz'}{{\footnotesize $-p_1{\bar z}'$}}
\psfrag{-q2}{{\footnotesize $-q_2$}}
\psfrag{x1p2}{{\footnotesize $x_1p_2$}}
\psfrag{x2p2}{{\footnotesize $x_2p_2$}}
\psfrag{p2}{{\footnotesize $p_2$}}
\psfrag{p2'}{{\footnotesize $p_2'$}}
\psfrag{kpx1p2}{{\footnotesize $k_\perp, x_1$}}
\psfrag{kpx2p2}{{\footnotesize $k_\perp, x_2$}}
\psfrag{pi}{{\footnotesize $\pi$}}
\psfrag{aa}{{\footnotesize $(a)$}}
\psfrag{bb}{{\footnotesize $(b)$}}
\psfrag{cc}{{\footnotesize $(c)$}}
\psfrag{dd}{{\footnotesize $(d)$}}
\psfrag{ee}{{\footnotesize $(e)$}}
\psfrag{ff}{{\footnotesize $(f)$}}
\psfrag{gg}{{\footnotesize $(g)$}}
\psfrag{Fipi}{{\footnotesize $\phi_\pi(z')$}}
\psfrag{Fchi}{{\footnotesize $F_\xi(x_1)$}}

\psfrag{zz}{{$z$}}
\psfrag{zeta}{{$\zeta$}}
\psfrag{TH}{\footnotesize $T_H(z',x_1)$}

\newcommand{\beq}[1]{
\begin{equation}\label{#1}}
\newcommand{\eeq}{\end{equation}}
\newcommand{\bea}[1]{
\begin{eqnarray}\label{#1}}
\newcommand{\eea}{\end{eqnarray}}

\begin{document}

\title{QCD Factorization for the
             Pion Diffractive
             Dissociation Into Two
             Jets}
\author{D.~Yu. Ivanov${}^{1,2}$}

\address{
${}^{1}$Institut f\"ur Theoretische Physik, Universit\"at Regensburg,\\
D-93040 Regensburg, Germany, \\
${}^{2}$Institute of Mathematics, 630090 Novosibirsk, Russia,\\
E-mail: Dmitri.Ivanov@physik.uni-regensburg.de}


\maketitle

\abstracts{
We report a detailed study of the process of  
pion
diffraction dissociation into two jets with
large transverse momenta. 
We find that the standard collinear
factorization does not hold in this reaction. 
The structure of non-factorizable contributions is discussed
and the results are compared with the
   experimental data. Our conclusion is that the existing theoretical
uncertainties do not allow, for the time
   being, for a quantitative extraction of the pion distribution amplitude.
(Talk presented at the Workshop on Exclusive Processes at High Momentum
Transfer, Jefferson News, VA, May 15-18, 2002)}

%\section{Introduction}
\noindent
{\bf 1.}
To our knowledge, the pion (and photon) diffraction dissociation into a
pair of jets with large transverse momentum on a nucleon target
was first discussed in \cite{KDR80}. Then the possibility
to use this process to probe the nuclear filtering of pion components
with a small transverse size was suggested in \cite{BBGG81}. 
The A-dependence and the
$q_\perp^2$-dependence of the
coherent dijet cross section was first calculated
in \cite{FMS93}. In the same work it was argued that the jet
distribution with respect to  the longitudinal momentum fraction has to
follow the quark momentum distribution in the pion and hence provides
a direct measurement
of the pion distribution amplitude. Recent experimental data
by the E791 collaboration \cite{E791a,E791b} indeed confirm the strong
A-dependence which is  a signature for color transparency, and are
consistent
with the predicted $\sim 1/q_\perp^8$
dependence on the jet transverse momentum. Moreover, the jet longitudinal
momentum fraction distribution turns out to be consistent with the
$\sim z^2(1-z)^2$ shape corresponding to the asymptotic pion distribution
amplitude which is also supported by an independent measurement of the pion
transition form factor $\pi\gamma\gamma^*$ \cite{CLEO}.
  
After these first successes, one naturally asks whether the
QCD description of coherent dijet production can be made fully quantitative.
Two recent studies \cite{NSS99} and \cite{FMS00} address this question,
with contradictory conclusions. Therefore we attempt to clarify
the situation and develop a perturbative QCD framework for the description
of coherent dijet production that would be in line with other known
applications of the QCD factorization techniques.
The results reported here have been obtained in collaboration with 
V.M. Braun, A. Sch\"afer and L. Szymanowski \cite{BISS01,Braun:2002wu}.

\begin{figure}[th]
%\epsfxsize=10cm   %width of figure - will enlarge/reduce the figures
%\epsfbox{fig3.eps}
%\figurebox{2cm}{3cm}{} %to have a box alone 
\centerline{\epsfxsize7.0cm\epsffile{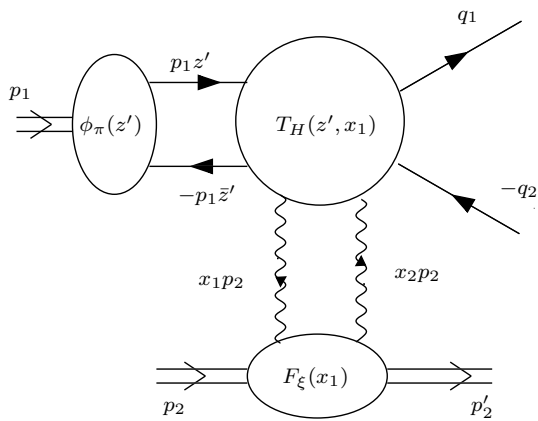}}
%\centerline{\epsfxsize=3.5in\epsfbox{procs-fig1.eps}}   
\caption{
Kinematics of the coherent hard dijet production $\pi\to 2\, {\rm jets}$.
The hard scattering amplitude $T_H$ contains at least one hard gluon
exchange. \label{fig:1}}
\end{figure}

\noindent
{\bf 2.}
The kinematics of the process is shown in Fig.~\ref{fig:1}.
%We consider $\pi^-$ scattering from the proton target.
The momenta of the incoming pion, incoming nucleon and the outgoing nucleon
are $p_1, p_2$ and $p_2^\prime$, respectively.
The pion and the nucleon masses are both neglected,
$p_1^2=0$, $p_2^2=(p_2^\prime)^2=0$.
We denote the momenta of the outgoing quark and antiquark (jets)
as $q_1$ and $q_2$, respectively. They are on the mass shell,
$q_1^2=q_2^2=0$, and can be decomposed
\beq{sudakov}
q_1=zp_1+\frac{q_{1\perp}^2}{zs}p_2+q_{1\perp} \ ,
\, q_2=\bar zp_1+\frac{q_{2\perp}^2}{\bar zs}p_2+q_{2\perp}
\eeq
such that $z$ is the longitudinal momentum fraction of the quark
jet in the lab frame.
We use the shorthand notation: $\bar u \equiv (1-u)$ for any
longitudinal momentum fraction $u$.
%The Dirac spinors for the quark and the antiquark are denoted by
%$\bar u(q_1)$ and $v(q_2)$.

We are interested in the forward limit, when the transferred momentum
$t=(p_2-p_2^\prime)^2$ is equal to zero
%\footnote{If the target mass $m$ is taken into account, the momentum
%transfer $t=(p_2-p_2^\prime)^2$ contains a non-vanishing longitudinal
%contribution and is constrained from below by
%$|t|\geq t_0$, where $t_0=({\displaystyle m^2M^4})/{\displaystyle
%(s-m^2)^2}$,
%$M^2$ being the invariant mass of the dijet.},
and the transverse momenta of jets compensate each other
$q_{1\perp} \equiv q_{\perp}$, $q_{2\perp}\equiv -q_\perp$.
In this kinematics the invariant mass of the produced $q\bar q$ pair is
equal to
$
M^2={\displaystyle q_{\perp}^2}/{\displaystyle z\bar z}
$.
The invariant c.m. energy $s=(p_1+p_2)^2=2p_1 p_2$
is taken to be much larger than the transverse jet momentum
$q_{\perp}^2$. 

\noindent
{\bf 3.} 
The possibility to constrain the pion distribution amplitude 
$\phi_\pi (z^\prime,\mu_F^2)$ in the dijet diffractive dissociation 
experiment assumes that 
the amplitude of this process can be calculated in the collinear
approximation as
suggested by  Fig.~\ref{fig:1}:
\beq{factor}
{\it M}_{\pi \to 2\, {\rm jets}}
=\sum_{p=q,\bar q, g} \int\limits^1_0 dz^\prime \int\limits^1_0 dx_1
\,\phi_\pi
(z^\prime,\mu_F^2)\,T^p_H(z^\prime , x_1, \mu_F^2)\,{\it F}^p_\zeta(x_1,
\mu_F^2)\,.
\eeq     
Here $F^p_\zeta(x_1, \mu_F^2)$ is the generalized parton
distribution (GPD) $p=q,\bar q,g$ 
\cite{early_skewed,Rad96a,Ji97a} in the target nucleon;
$x_1$ and $x_2=x_1-\zeta$ are the 
momentum fractions of the emitted and the absorbed partons, respectively.
The asymmetry parameter $\zeta$ is fixed by the process kinematics:
$\zeta = M^2/s ={\displaystyle q_{\perp}^2}/{\displaystyle z\bar z}$.
$T_H(z^\prime , x_1, \mu_F^2)$ is the hard scattering amplitude involving at
least one hard gluon exchange and $\mu_F$ is the (collinear) factorization
scale. By definition, the pion distribution amplitude only involves
small momenta, $k_\perp < \mu_F$, and the hard scattering amplitude
is calculated  neglecting the parton transverse
momenta. 

We calculated \cite{Braun:2002wu} both the leading-order 
gluon and quark contributions to
the amplitude 
and find that in both cases the corresponding hard kernels $T^q_H$, $T^g_H$
diverge as $1/z'^2$ and $1/\bar z'^2$ in the $z'\to 0$ and $z'\to 1$ limit,
respectively. This implies that the integration of the pion momentum
fraction diverges at the end-points and the collinear factorization is,
therefore, broken. 

At high energies the contribution of gluon GPD dominates. We found that
up to kinematical factors
\bea{G-coefffun}
T_{H}^g &= &C_F
\left(
\frac{\bar z}{z^\prime}+\frac{z}{\bar z^\prime}  
\right)
\left(
\frac{\zeta}{[x_1-i\epsilon]^2}+
\frac{\zeta}{[x_2+i\epsilon]^2} -
\frac{\zeta}{[x_1-i\epsilon][x_2+i\epsilon ]}
\right) \nonumber \\
&&{}
+
\left(
\frac{z\bar z}{z'\bar z'}+1
\right)
\left[
C_F
\left(
\frac{z\bar z}{z'\bar z'}+1
\right)
+\frac{1}{2N_c}
\left(
\frac{z}{z'}
+\frac{\bar z}{\bar z'}
\right)    
\right]    
\nonumber \\
&&{}
\times
\left(           
\frac{1}{[(z-z^\prime)x_1-z\bar z^\prime \zeta+i\epsilon ] }
+                
\frac{1}{[(z^\prime -z)x_2-z\bar z^\prime \zeta+i\epsilon ]}
\right)          
  \\             
&&{}             
-                
\left[           
C_F              
\frac{z\bar z}{z'\bar z'}
\left(\frac{\bar z}{z'}+\frac{z}{\bar z'}\right)
+\frac{1}{2N_c\,z' \bar z'}\left(\frac{z\bar z}{z'\bar z'}+1\right)
\right]          
\frac{\zeta}{[x_1+i\epsilon ][x_2-i\epsilon ]}  \ . \nonumber
\eea  
The differential cross section summed over the
polarizations and the color of quark jets is given by
\beq{cross} 
\frac{d\sigma_{\pi\to 2\,{\rm jets}}}{d q_\perp^2 dt dz}
= \frac{\alpha_s^4 f_\pi^2 \pi(1-\zeta)}{8N_c^3 q_\perp^8} |{\it M}|^2\, ,
\label{crosection}
\end{equation}
where ${\it M}$ is 
calculated as in (\ref{factor}) with $T_{H}^g$ given in (\ref{G-coefffun}), 
$f_\pi =133$ MeV is the
pion decay constant.  

\noindent
{\bf 4.} 
According to Eq.~(\ref{G-coefffun}) 
the leading end-point behavior of the gluon amplitude 
at $z'\to 0$ and
$z'\to 1$ is given by the following expression
\beq{end}
 {\it  M}\Big|_{\rm end-points}
  = -i\pi \left(N_c+\frac{1}{N_c}\right)
 z \bar z \,  \int\limits^1_0 dz'\,
\frac{\phi_\pi(z',\mu^2)}{z'^2}{\it F}_\zeta^g(\zeta,\mu^2) \,.
\eeq
Since $\phi_\pi(z')\sim z'$ at $z'\to 0$, the integral over $z'$
diverges logarithmically.
Remarkably, the integral containing the pion distribution
amplitude does not involve any $z$-dependence.
Therefore, the longitudinal momentum distribution of the jets in the
nonfactorizable contribution is calculable and, as it turns out,
has the shape of the asymptotic pion distribution amplitude
$\phi_\pi^{\rm as}(z) = 6z\bar z$.  

In technical terms, the appearance of the end point divergency
is due to pinching of the $x_1$ contour in the point $x_1=\zeta (x_2=0)$ 
in the case when variable $z^\prime$
is closed to the end points, cf. Eqs.~(\ref{factor},\ref{G-coefffun}). 
One can trace \cite{Braun:2002wu} that this pinching occurs 
between soft gluon (the gluon with momentum $x_2\to 0$) 
interactions in the initial and in the final state, and is related with 
the existence of the unitarity cuts of the amplitude in different, 
$s$ and $M^2$, channels.

The other important integration region in (\ref{factor}) is the one when  
$\zeta \ll |z'- z| \ll 1$, i.e. 
when the longitudinal momentum fraction carried by the quark is close
(for high energies) to that of the quark jet in the final state
%The enhancement of this region comes from the (small)
%denominator $1/(z'-z)$ which  is present in the contributions in
%Fig.~\ref{fig:3}c,d with real gluon emission in the intermediate
%state. 
\bea{z=z'}
{\it M}\Big|_{\zeta\ll |z'- z|\ll 1}
& = & -4 i\pi N_c \,\phi_\pi(z)\,\int\limits^{1}_{z}
  \frac{dz'}{z'-z} {\it F}_\zeta^g(\zeta
\frac{z\bar z}{z'-z},q^2_{\perp}) \nonumber \\
& \simeq & -4 i \pi N_c \,\phi_\pi(z)\!\int\limits_\zeta^1
 \!\frac{dy}{y}\, {\it F}_\zeta^g(y,q^2_{\perp})\,.
\eea
This logarithmic integral is nothing but
the usual energy logarithm that accompanies each extra gluon in the
gluon ladder. Its appearance is due to the fact the hard gluon
which supplies jets by the high transverse momentum 
can be emitted in a broad rapidity interval and
is not constrained to the pion fragmentation region.
The integral on the r.h.s. of (\ref{z=z'}) can be identified with the
unintegrated generalized gluon distribution. 
And, therefore, in the region $z'\sim z$ hard
gluon exchange can be viewed as a large transverse momentum part of the
gluon distribution in the proton, cf. \cite{NSS99}.
This contribution is proportional to
the pion distribution amplitude $\phi_\pi (z,q^2_{\perp})$
and contains the enhancement factor $\ln 1/\zeta \sim \ln s/q_\perp^2$.

%Physically this means that the approximation of
%neglecting
%the incident quark transverse momenta becomes insufficient close to the
%end points.

\begin{figure}[hbtp]
\centerline{\epsfxsize8.0cm\epsffile{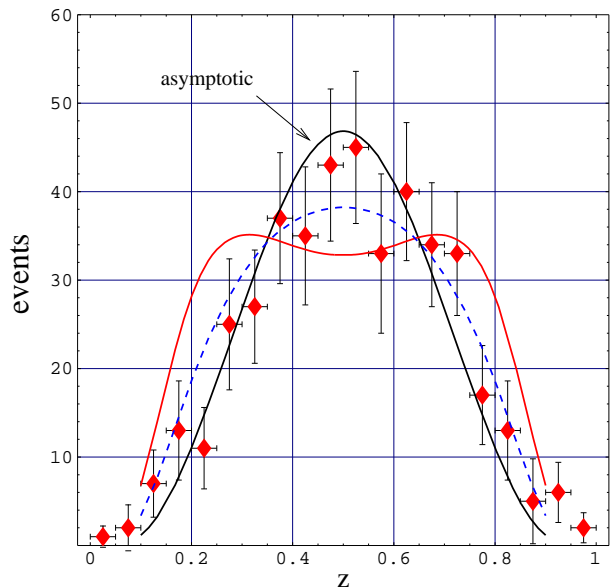}}
\caption[]{\small 
The longitudinal momentum fraction distribution of the dijets
%with $1.5 \le q_\perp \le 2.5$~GeV 
from the platinum target \cite{E791a} 
in comparison with our predictions, see text.
%The two solid curves show the calculations with the two extreme
%pion distribution
%amplitudes  --- asymptotic  and `two-humped'',
%respectively. The dashed curve corresponds to the Chernyak-Zhitnitsky
%model evolved to the scale $\mu = 2$~GeV.
%The overall normalization (the same for all curves) is arbitrary.
 }
\label{E791}
\end{figure}

\noindent
{\bf 5.} 
We performed numerical calculations for the kinematics of 
E791 experiment: the transverse momentum range
$1.5 \le q_\perp \le 2.5$~GeV and $s=1000$~GeV$^2$.
We found that the  diffractive (\ref{z=z'}) and the end-point (\ref{end}) 
contributions are numerically comparable to each other. In the later case 
in order to regularize the end-point divergence
we used the simplest prescription,  an explicit cutoff
on the quark momentum
fraction in the pion $z^\prime\geq \mu_{\rm
IR}^2/q^2_\perp$, where $\mu_{\rm IR}$ is of order of 
intrinsic quark transverse momentum in the pion, see \cite{Braun:2002wu} 
for more details. For GPD's the parametrization \cite{GPDs} was used.
Fig.~\ref{E791} shows the comparison
of the calculated dijet momentum fraction distribution
with the data \cite{E791a}. 
The two solid curves correspond to the asymptotic, $\phi_\pi^{\rm as}(z)$, 
and `two-humped'',
$\phi_\pi^{\rm CZ}(z,\mu = 0.5~\mbox{\rm GeV})  =
   30 z(1-z)(1-2z)^2$, forms of the pion distribution amplitude. 
The dashed curve corresponds to the Chernyak-Zhitnitsky   
model evolved to the scale $\mu = 2$~GeV, $\phi_\pi^{\rm CZ}(z,\mu =
2~\mbox{\rm GeV})  =
   15 z(1-z)[0.20+(1-2z)^2$.
The overall normalization is arbitrary, but is the same for all three
choices of the distribution amplitude.  
It is seen that experimental 
uncertainties do not allow for the separation between the 
distribution amplitudes $\phi_\pi^{\rm as}(z)$ and $\phi_\pi^{\rm CZ}(z,\mu
=2~\mbox{\rm GeV})$ while the extreme
choice $\phi_\pi^{\rm CZ}(z,\mu = 0.5~\mbox{\rm GeV})$ is not favored.  
This general conclusion is in agreement with the analysis
in \cite{CG01}.

%In what follows we often neglect
%contributions to the amplitude that are suppressed by powers of
%$q_\perp^2/s$.
%
%%%%%%%%%%%%%%%%%%     FIGURE E791  %%%%%%%%%%%%%%%%%%%%%%%
%%%%%%%%%%%%%%%%%%%%%%%%%%%%%%%%%%%%%%%%%%%%%%%%%%%%%%%%%%%%%%%%%%%%%%
%

\noindent
{\bf 6.} 
Our calculation is 
close in spirit to \cite{Che01,CG01} although the conclusion about 
the factorization is different.    
We examined the transition to the light-cone limit
carefully and found that the approximation used in \cite{Che01,CG01}
breaks down for soft gluons (and quarks). 
In the double 
logarithmic approximation our result in (\ref{z=z'}) is similar to
\cite{NSS99} obtained using different methods. We, therefore,
agree with the interpretation suggested in \cite{NSS99} that in the
true diffraction limit, for very large energies, the dijet production
can be considered as a probe of the hard component of the pomeron.
We note, however, that this interpretation breaks down beyond the  
double logarithmic approximation and is not sufficient
for the energy region of the E791 experiment. Finally, we have to   
mention an approach to coherent diffraction suggested 
in \cite{FMS00} that attributes hard dijets to a hard component
of the pion wave function. 
This technique is interesting, 
but apparently complicated
for the discussion of factorization. We found that the general 
argumentation in \cite{FMS00}
appears to be in contradiction with our explicit calculations.

\section*{Acknowledgments}
This work was supported by Alexander von Humboldt Foundation.

\end{document}